\documentstyle[12pt,emulateapj]{article}

\slugcomment{Submitted to PASP}

\begin{document}

\title{The Type IIn Supernova 2002kg: The Outburst of a Luminous
Blue Variable Star in NGC 2403\footnote{Based in part on observations
with the NASA/ESA Hubble Space Telescope, obtained at the Space
Telescope Science Institute (STScI), which is operated by AURA, Inc.,
under NASA contract NAS5-26555.}}

\author{Schuyler D.~Van Dyk}
\affil{Spitzer Science Center/Caltech, Mailcode 220-6, Pasadena CA  91125}
\authoremail{vandyk@ipac.caltech.edu}

\author{Weidong Li and Alexei V.~Filippenko}
\affil{Department of Astronomy, 601 Campbell Hall, University of
California, Berkeley, CA  94720-3411}
\authoremail{weidong@astro.berkeley.edu, alex@astro.berkeley.edu}

\author{Roberta M. Humphreys}
\affil{Department of Astronomy, 116 Church St. SE, University of
Minnesota, Minneapolis, MN  55455}
\authoremail{roberta@isis.spa.umn.edu}

\author{Ryan Chornock and Ryan J. Foley}
\affil{Department of Astronomy, 601 Campbell Hall, University of
California, Berkeley, CA  94720-3411}
\authoremail{chornock@astro.berkeley.edu, foley@astro.berkeley.edu}

\and

\author{Peter M. Challis}
\affil{Harvard-Smithsonian Center for Astrophysics, 60 Garden St.,
Cambridge, MA 02138}
\authoremail{pchallis@cfa.harvard.edu}

\begin{abstract}
We show that Supernova (SN) 2002kg in NGC 2403, initially classified
as Type II-narrow (IIn), has photometric and spectroscopic properties
unlike those of normal SNe.  Its behavior, instead, is more typical of
highly massive stars which experience the short-lived luminous blue
variable (LBV) phase toward the end of their lives.  The star, in
fact, most resembles the LBV S Doradus in outburst.  The precursor of
SN 2002kg is the irregular, bright blue variable star 37 (V37),
catalogued by Tammann \& Sandage in 1968.  Using high-quality
ground-based, multi-band images we can constrain the initial mass of
V37 to be $M_{\rm ini} \gtrsim 40\ M_{\odot}$.  We find that, although
the spectra indicate a nitrogen enhancement, possibly revealing the
products of CNO processing by V37 in the ejecta, the star lacks a
substantial LBV nebula.  The outburst from SN 2002kg/V37 is not nearly
as energetic as the giant eruptions of the $\eta$ Carinae-like
variables, such as SN 1954J/V12, also in NGC 2403.  SN 2002kg/V37,
however, is among a growing number of ``SN impostors'' exhibiting a
broad range of outburst energetics during a pre-SN phase of
massive-star evolution.
\end{abstract}


\section{Introduction}

The evolution of the most massive stars is not well understood.  After
the main sequence, stars with masses $M \gtrsim 20$--$30\ M_{\sun}$
should go through the red supergiant phase, or directly to a
short-lived luminous blue variable (LBV) phase, to become Wolf-Rayet
stars (e.g., Langer et al.~1994; Stothers \& Chin 1996) before
termination as supernovae (SNe; e.g., Woosley, Langer, \& Weaver 1993)
or ``collapsars'' (e.g., MacFadyen \& Woosley 1999).  LBVs which
experience normal eruptions or outbursts generally increase in
brightness by one or two visual magnitudes, but remain relatively
constant in bolometric luminosity (Humphreys \& Davidson 1994).
However, the extremely rare case of $\eta$ Carinae (e.g., Davidson \&
Humphreys 1997) 
shows
that some very massive stars go through tremendous eruptive phases of
mass ejection prior to the end of their lives.  
An estimate of the initial mass for $\eta$ Car is $\gtrsim$150 $M_{\sun}$ 
(Hillier et al.~2001).
The very massive stars define the upper luminosity
boundary of the Hertzsprung-Russell (HR) diagram.  It would, of
course, be very instructive to determine the prevalence of (and mass
range for) the stars which undergo outbursts prior to becoming a SN.

Several recent luminous events which have been identified as SNe are
probably not genuine SNe at all.  Instead, it is argued that these are
more likely to be super-outbursts of very massive stars, analogous to
$\eta$ Car (Goodrich et al.~1989; Filippenko et al.~1995; Van Dyk et
al.~2000).  The energetics are comparable to those of SNe; hence,
these objects act as SN ``impostors.''  The ``classical'' examples are
SN 1954J/V12 in NGC 2403 (Humphreys, Davidson, \& Smith 1999; Smith,
Humphreys, \& Gehrz 2001; Van Dyk et al.~2005) and SN 1961V in NGC
1058 (e.g., Van Dyk, Filippenko, \& Li 2002, Van Dyk 2005; however, see
Chu et al.~2004).  Two more well-studied cases are SN 1997bs in M66
(Van Dyk et al.~2000) and SN 2000ch in NGC 3432 (Wagner et al.~2004).
These objects were all underluminous and exhibited photometric
behavior unlike that of any normal SNe.  For SN 1997bs, Van Dyk et al.~(1999)
identified a luminous precursor star, with ${M_V}_0 \simeq -8.1$ mag,
in an archival {\sl Hubble Space Telescope \/} ({\sl HST}) WFPC2 F606W
image.  Unfortunately, no color information was available for the star
to compare it to theoretical evolutionary models, in order to derive
or strictly constrain the star's mass.

In this paper we present another similar case, SN 2002kg, also in NGC
2403, which we similarly argue is not actually an explosion at the end
of a massive star's life, but instead is the powerful eruption of a
LBV.  SN 2002kg was discovered on 2003 October 22 (UT dates are used
throughout this paper) 
by Schwartz \& Li (2003) with the Tenegra II
and Katzman Automatic Imaging Telescope (KAIT; Li et al.~2000;
Filippenko et al.~2001; Filippenko 2005); it first became visible in
images taken on 2002 October 26 (UT dates are used throughout this
paper).

In Figure 1 we show a KAIT
image of the SN in the host galaxy.  Filippenko \& Chornock (2003)
identified the SN spectroscopically as a Type II-``narrow'' supernova
(SN~IIn; Schlegel 1990, Filippenko 1991, 1997); it has Balmer emission
lines exhibiting a narrow component atop a broader base, but with
characteristics more similar to those of SN 1997bs (Van Dyk et
al.~2000) than to the prototypical SNe~IIn (e.g., SN 1988Z, Turatto et
al.~1993; SN 1995N, Fransson et al. 2002).  In Van Dyk (2005) we first
identified SN 2002kg with a previously known luminous irregular blue
variable (V37) in the host galaxy and provided an estimate of the initial
mass of this precursor star.  Here we present a more detailed
discussion.  Weis \& Bomans (2005) also have recently identified SN
2002kg with V37 and discuss its historical light curve.

\section{Observations}

\subsection{Photometry of the Outburst}

\subsubsection{KAIT Monitoring}

Since the host galaxy has been monitored frequently by KAIT we are
able to obtain unfiltered (${\sim}R$) photometry for the SN 2002kg
outburst over a period of more than one year.  Li et al.~(2003)
discuss how the unfiltered magnitudes for an object can be transformed
to standard $R$ magnitudes, with knowledge of its detailed color
evolution.  If the object's color remains unchanged, then the
transformation is merely a constant offset.  We find that the SN has
approximately the same color (within $\sim$0.1 mag) at both an early
and late time in the outburst (see \S 5); we could therefore follow
the prescription in Li et al.~(2003) to derive the offset.

Equivalently, we can simply adjust the KAIT magnitudes, by about
$-0.7$ mag, based on the offset between the unfiltered magnitude from
the March 25.17 (JD 2452723.67) KAIT observation and the $R$-band
magnitude from the March 26.15 (JD 2452724.65) Palomar observation
(see \S 2.1.2; we consider the likely change in the $R$ magnitude over
about one day to be insignificant for our purposes).  We note that
systematic errors in the magnitudes probably still exist for this KAIT
light curve, due to the broad spectral range of the unfiltered images
and the assumption of constant color for SN 2002kg (which may not apply
for the entire outburst).  We list the resulting $R$ magnitudes in
Table 1 and show the light curve in Figure 2, after further correcting
for both extinction (see \S 3) and the true distance modulus to NGC
2403 ($\mu_0=27.48$ mag; Freedman et al.~2001).

The light curve shows highly erratic and unusual behavior for a SN.
The maximum $R$ luminosity is about $-10.2$ mag, more than 7 mag
fainter than a typical SN IIn (see Van Dyk et al.~2000, their Figure
5).  In fact, this light curve is more qualitatively similar to those
of the possible $\eta$ Car analog, SN 2000ch (Wagner et al.~2004,
their Figure 3), or the LBV S Doradus (Humphreys \& Davidson 1994;
their Figure 2).  Schwartz \& Li (2003) note that SN 2002kg is not
detectable, $R \gtrsim 19.5$ mag ($M_R \gtrsim -8.0$ mag), in a KAIT
image from 1998 November 13.  Apparently we first caught the SN with
KAIT while it was already undergoing the observed outburst (this is
supported by the lack of detection at earlier epochs, based on data
from other telescopes; see \S 2.1.2).  The absolute magnitude of this
outburst was already $\gtrsim$2 mag brighter at that time than when
the precursor star was in quiescence (see also \S 3).

As Schwartz \& Li (2003) point out, the light curve shows a
brightening trend between 2002 October 26, when the outburst was first
noticed, and 2003 early January.  The light curve then shows a
noticeable dip after 2003 January 1, before reaching maximum near 2003
mid-March.  The brightness declines yet again by late March.  No
coverage exists from KAIT between 2003 late March and mid-October, so
we have no knowledge of the outburst behavior during that interval.
However, by October 11 SN 2002kg has approximately the same brightness as
it did in late March.  During 2003 October and November the object
remains at a relatively constant brightness. We then see a pronounced
dip in the light curve, after 2003 late November through 2004 early
March.  The light curve rises again into 2004 early April, at which
time our KAIT coverage terminates.  This is similar to the behavior of
SN 2000ch; Wagner et al.~(2004) speculate that the dip seen for SN
2000ch may be due to either a dust formation event, an eclipse, or
changes within an optically thick ejected envelope.  For SN 2002kg,
this dip is much broader in time than was seen for SN 2000ch.  We
suspect that for SN 2002kg the light curve dip is more likely due to
opacity changes in the ejected envelope than to dust formation (see \S 4).

\subsubsection{Other Ground-Based Imaging}

As already mentioned, we obtained $R$, as well as $BVI$, images of 
SN 2002kg with the Palomar 1.5-m telescope on 2003 March 26.15.  In Figure 3
we show the $V$-band image.  We extracted instrumental magnitudes for
the relatively bright SN 2002kg in the Palomar images via fitting an
empirically derived model point-spread function (PSF) for each band,
using DAOPHOT (Stetson 1987) within IRAF\footnote{IRAF (Image
Reduction and Analysis Facility) is distributed by the National
Optical Astronomy Observatories, which are operated by the Association
of Universities for Research in Astronomy, Inc., under cooperative
agreement with the National Science Foundation.}.  Unfortunately, no
photometric calibration was obtained during these Palomar
observations.  However, Larsen \& Richtler (1999) also obtained
$UBVRI$ (plus H$\alpha$) images of NGC 2403 with the Nordic Optical
Telescope (NOT)\footnote{The NOT images were contributed to the
NASA/IPAC Extragalactic Database (NED) and posted for public
distribution on their website, http://nedwww.ipac.caltech.edu.} on
1997 October 13, prior to SN 2002kg, under very good seeing conditions
of $0{\farcs}8$.  (We will discuss these NOT images in more detail in
\S 3.)  We extracted instrumental magnitudes from the NOT broad-band
images using IRAF, first with a 4\arcsec\ aperture and then with a PSF
for each band.  Calibration was established using photometry of a
number of isolated stars through a 20\arcsec\ aperture, which matches
the aperture used to establish the calibration through standard-star
observations by Larsen (1999; we also applied the appropriate
extinction corrections, supplied via a private communication from
S.~S.~Larsen).  We have compared our photometry with the $BV$
bright-star photometry of Sandage (1984) and $UBVR$ photometry of
Zickgraf \& Humphreys (1991) and find good agreement across all bands.

We tie the photometry from the Palomar images to that from the
calibrated NOT images and list the results in Table 2.  As mentioned
above in \S 2.1.1, we used the $R$ magnitude for SN 2002kg from these
Palomar data to adjust the unfiltered KAIT light curve.  We also
include this $R$ magnitude from Palomar in the light curve shown in
Figure 2.

A number of images of NGC 2403 obtained in several bands prior to 2002
October are also available in the Isaac Newton Group archive.  These
images have a range of sensitivities, and SN 2002kg is not detected in
any of them.  We list the resulting upper limits to detection of the
outburst in Table 2.

\subsubsection{{\sl HST\/} Imaging}

Additionally, SN 2002kg was observed on 2004 August 17 with the Wide
Field Channel (WFC) and on 2004 September 21 with the High Resolution
Channel (HRC) of the Advanced Camera for Surveys (ACS), on-board {\sl
HST}.  The August observations were obtained during a campaign by
program GO-10182 (PI: Filippenko) to observe SN 2004dj, the third
historical SN in NGC 2403.  The imaging strategy for SN 2004dj was
purposely designed so that the site of SN 2002kg would also be
contained in the WFC images.  The September observations were obtained
as part of a larger Snapshot program observing SNe (GO-10272; PI:
Filippenko).  The WFC exposures were 600 s in F475W, 650 s in F658N,
and 350 s each in F606W and F814W.  The HRC exposures were 840 s in
F435W and 360 s in F625W.  The images were all extracted from the {\sl
HST\/} archive.  For the WFC data we recombined the individual ``flt''
exposures in all four bands at the Space Telescope Science Institute
(STScI), in order to more reliably remove cosmic-ray hit features from
the images, and, in effect, produce our own drizzled (Fruchter \& Hook
2002) ``drz'' images independent of the {\sl HST\/} data pipeline.

The WFC F606W image is shown in Figure 4.  We performed PSF-fitting
photometry within IRAF/DAOPHOT on the broad-band WFC images.  The
empirical PSF for each band was derived from the $\sim 10$ brightest,
uncrowded, and unsaturated stars in these images, based on a
$0{\farcs}5$ aperture.  For the WFC narrow-band image, we performed
aperture photometry only, using a $0{\farcs}5$ aperture.  We
determined that any correction to the magnitudes for the degradation
of charge-transfer efficiency (CTE) is $<$0.2\% in all bands (see
Riess \& Mack 2004).  Next, for the F475W and F606W filters we applied
the corrections from a $0{\farcs}5$ aperture to infinite aperture from
Table 5 in Sirianni et al.~(2005); for the F814W filter, we applied
the correction from their Table 6, assuming the effective wavelength
from their Table 8 for approximate spectral type F2.  We then applied
the zero points (VEGAMAG for the broad bands, and STMAG for the narrow
band) in Table 10 of Sirianni et al.~to the corrected instrumental
magnitudes.  The resulting flight system magnitudes are $m_{\rm
F475W}=19.43 \pm 0.02$, $m_{\rm F606W}=19.26 \pm 0.02$, $m_{\rm
F814W}=19.13 \pm 0.03$, and $m_{\rm F658N}=18.67 \pm 0.14$.
 
In the HRC images SN 2002kg is relatively isolated, so we simply
employed aperture photometry (assuming a $0{\farcs}5$ radius aperture)
within IRAF.  We then applied the corrections to infinite aperture
from Table 5 and the zero points from Table 11 in Sirianni et
al.~(2005).  The resulting flight system magnitudes are $m_{\rm
F435W}=19.41 \pm 0.05$ and $m_{\rm F625W}=19.16 \pm 0.04$.

Finally, we transformed the broad-band magnitudes into the
Johnson-Cousins system, also following Sirianni et al., keeping in
mind the fact that SN 2002kg shows strong emission lines (Filippenko
\& Chornock 2003, and also \S 2.2), whereas the stars used to
establish the transformations all show normal absorption-line spectra.
We list the resulting magnitudes in the standard photometric system in
Table 2.  We also include the HRC $R$-band data in the light curve
shown in Figure 2.

\subsection{Spectroscopy of the Outburst}

Optical spectra of SN 2002kg were obtained at the Keck 10-m telescope
with the Low Resolution Imaging Spectrometer (LRIS; Oke et al. 1995)
on 2003 January 6 (the confirmation spectrum; JD 2452646.0) during the
early brightening trend, on 2003 November 29 (JD 2452973.0),
prior to the pronounced dip in brightness, and on 2004 December 12
(JD 2453352.03) at relatively later times.  These observations had
spectral resolution of $\sim$6 \AA\ across most of the spectral range,
3000--9430 \AA\ (the resolution is somewhat worse in the
near-infrared).  The 2004 December spectrum was obtained with a
600-line grism on the blue side, whereas the other two LRIS spectra
were obtained with a 400-line grism in the blue.
At H$\alpha$ the resolution (FWHM) of these spectra is
$\sim$320 km s$^{-1}$.  

In addition, a spectrum was obtained on 2003 November 20 (2452964.0),
while the object was still bright, with the blue channel of the
spectrograph on the 6.5-m Multiple-Mirror Telescope (MMT). A 300-line
grating was used, with spectral range 3700--8300 \AA.   
For the Keck spectra conventional data reductions
were performed within IRAF, including bias subtraction, flat-fielding,
and wavelength calibration, while our own IDL\footnote{IDL is the
Interactive Data Language, a product of Research Systems, Inc.}
routines were used for flux calibration and removal of telluric lines
(e.g., Matheson et al. 2000). Reduction of the MMT spectrum was
performed in a similar fashion.

The spectra are shown in Figure 5.  The overall shape of the continuum
has remained relatively blue throughout 2003 and 2004, peaking near
3700 \AA.  The spectra all show several strong emission lines,
primarily of H, but also of possible weak Fe II at 4500--4600 \AA\ and
at 5150--5350 \AA, as well as possible blends of H with He I.  Figure
6 shows just the H$\alpha$ plus [N~II] $\lambda\lambda$6548, 6584~\AA\
lines.  The narrow Balmer profiles are unresolved, and the broad
profiles at H$\alpha$ remain relatively constant throughout 2003 and
2004, with Gaussian velocity widths $\sigma \approx 370$ km s$^{-1}$.
Such a velocity width for the line is uncharacteristically low for a
young SN, but consistent with the terminal wind velocities for LBVs.
The presence of the prominent, unresolved [N~II] emission
lines, while other forbidden lines (e.g., due to O and S) appear weak
or absent, suggests a possible N enhancement in the circumstellar gas.

From the line profiles in Figure 6 it is clear that the power in the
H$\alpha$+[N~II] emission has steadily declined. This trend is
consistent with that of the $R$-band light curve for the observation
dates of the spectra.  In fact, integrating over the
continuum-subtracted H$\alpha$+[N~II] line profiles shown in
Figure 6, we find an observed flux $F_{{\rm H}\alpha+[{\rm N II}]}
\approx 1.3 \times 10^{-14}$ erg cm$^{-2}$ s$^{-1}$ (uncorrected for
extinction) in 2003 January, $\sim 1.1 \times 10^{-14}$ erg cm$^{-2}$
s$^{-1}$ in 2003 November, and $\sim 4.5 \times 10^{-15}$ erg
cm$^{-2}$ s$^{-1}$ in 2004 December.  (The absolute flux scale for the
Keck spectra is accurate to $\sim 20$\% or better.)

In the blue portion of the spectra, shown in Figure 7, we see most
notably the presence of various absorption lines, mostly due to
higher-order Balmer lines, and possible blends with He~I.  
Strong Ca~II H and K
lines (the H component is also probably blended with H$\epsilon$) are
also seen, particularly in 2003, but we suspect these lines are
dominated by interstellar absorption local to the SN (the extinction
toward the SN appears to be significant; see \S 3).

Such absorption features are not typically seen in SN spectra.  (We
have been very careful in the extraction of the spectra from the
2-dimensional images and can safely rule out contamination by a
neighboring star.)  The equivalent widths of these lines indicate an
underlying star with spectral type and luminosity class of a late O or
early-to-mid B supergiant.  We compare the SN 2002kg spectra to those
of the LBV S Dor (from 1996, during outburst; Massey 2000) and the
well-studied S Dor-type LBV AG Carinae (in a deep minimum state;
Walborn \& Fitzpatrick 2000).  Although the Balmer emission lines for
SN 2002kg lack the P-Cygni profiles seen in the AG Car spectrum, the
SN spectra, overall, show a remarkable resemblance to that for S Dor
in outburst.  We therefore consider the spectroscopic evidence quite
compelling that SN 2002kg is not a true SN, but instead a S Dor-like
LBV in outburst.  A similar conclusion was also reached by Weis \&
Bomans (2005).

\section{Variable 37: The Very Massive Precursor of SN 2002kg}

From the Palomar images (\S 2.1.2) we measure an accurate absolute
position for the SN of $\alpha$(J2000) = $7^h\ 37^m\ 01{\fs}66$,
$\delta$(J2000) = $65\arcdeg\ 34\arcmin\ 28{\farcs}0$, with a total
uncertainty of $\pm 0{\farcs}2$, using the Two Micron All Sky Survey
(2MASS) as the astrometric reference (see Van Dyk, Li, \& Filippenko
2003).  This position differs by $1{\farcs}7$ from that reported by
Schwartz \& Li (2003), and results in an offset from the host galaxy's
nucleus of $63{\farcs}6$ east and $101{\farcs}2$ south.  We then
applied the same astrometric grid to the NOT images (with a $\pm
0{\farcs}2$ total uncertainty in the grid).  In Figure 8 we show the
NOT $V$-band image from 1997.  When the absolute SN position is
superimposed on this image, it coincides {\it exactly\/} with the star
indicated by the arrow in the figure.  This star is almost certainly
the precursor of SN 2002kg.

We find that this star had $V=20.63 \pm 0.04$, $U-B=-0.96 \pm 0.05$,
$B-V=-0.08 \pm 0.05$, $V-R=0.09 \pm 0.04$, and $V-I=-0.06 \pm 0.05$
mag (see Table 2).  (Uncertainties include the uncertainties in the
standard calibration in each band from Larsen 1999, his Table 1, added
in quadrature with the measurement uncertainties from IRAF.)  The
fainter ($V=21.01 \pm 0.04$ mag), redder ($V-I=0.50 \pm 0.05$ mag)
star $\sim 1{\farcs}0$ to the northeast is well outside the total
uncertainty in the SN position in the NOT image, $0{\farcs}3$ (summing
in quadrature the total uncertainty in the SN position measured from
the Lick image and the uncertainty in the astrometric grid on the NOT
image).  Also, this fainter star appears at approximately this same
brightness (to within one magnitude) and color in the {\sl HST\/}
images as in the NOT images, whereas SN 2002kg (and therefore its
precursor) have changed more significantly both in brightness and in
color (compare Figure 8 with Figure 3 and see Table 2).  We note that
Weis \& Bomans (2004) also attempt to identify the precursor, but,
using pre-outburst ground-based images with inferior spatial
resolution, they are unable to distinguish the star we have identified
as the precursor from its fainter neighbor without consulting the
post-outburst {\sl HST\/} images.

This precursor star had actually been identified first by Tammann \&
Sandage (1968) in their photographic study of supergiants in NGC 2403.
We have determined this from a careful astrometric comparison of their
Figure 5, an enlargement of a 103aO photographic plate of the galaxy,
with the NOT $V$-band image.  The SN precursor star we have identified
lines up very well with the position of the catalogued bright blue
irregular variable star, no.~37 (hereafter, V37) from Tammann \&
Sandage (1968).  We consider the positional coincidence a compelling
indication that the two stars are one and the same (see also Weis \&
Bomans 2005).

Tammann \& Sandage (1968) describe the $B$ light curve for V37 as
exhibiting relatively small amplitudes of $\sim 2$ mag, between about
19 and 21 mag, and characteristic fluctuation timescales of several
years (see their Figure 10).  Where data were available between about
1948 and 1963, the observed $B-V$ color of V37 apparently fluctuated
between $+0.20$ and $-0.36$ mag.  The maximum absolute $B$ magnitude
for V37 during their monitoring was $\sim -8.8$, although generally it
was $\sim -6.5$ mag.

We can estimate the extinction and reddening to this star by placing
it in color-color diagrams (Figure 9), based on the NOT and {\sl
HST\/} ACS/WFC photometry.  Also shown in both diagrams are the loci
of the main sequence and supergiants (Drilling \& Landolt 2000), and
the reddening vector for OB supergiants, assuming a Cardelli, Clayton,
\& Mathis (1989) reddening law.  Figure 9a, in particular, implies
that the star is a blue supergiant reddened along this vector, at $A_V
= 0.6 \pm 0.2$ mag ($E[B-V] = 0.19 \pm 0.06$ mag).  Most of this
extinction must be in the star's local environment; the component of
the Galactic foreground extinction alone is only $A_V = 0.13$ mag
(Schlegel, Finkbeiner, \& Davis 1998).  In Figure 9 we also show the
loci of stars in the precursor's immediate ($\sim 200$ pc)
environment; the reddening in the environment appears to be quite
variable, with many stars more reddened than the precursor.  (The
fainter star $\sim 1\arcsec$ to the northeast is quite reddened, with
$A_V \gtrsim 2$ mag.)  The extinction toward the precursor is
generally consistent with the interstellar extinction for stars in its
environment.

Correcting the star's observed photometry for the implied extinction
and reddening, as well as for the true distance modulus of NGC 2403
($\mu_0 = 27.48 \pm 0.1$ mag; Freedman et al.~2001), we find
${M_V}^0=-7.45 \pm 0.23$, $(U-B)_0=-1.02 \pm 0.07$, $(B-V)_0=-0.13 \pm
0.08$, $(V-R)_0=-0.01 \pm 0.05$, and $(V-I)_0=-0.30 \pm 0.09$ mag.
(The total uncertainty in the absolute magnitude is the measurement
uncertainty and the uncertainty in the distance modulus, added in
quadrature; the total uncertainties in the colors include the
measurement uncertainties and the uncertainty in the reddening,
assuming the Cardelli et al.~1989 reddening law, again added in
quadrature.)  The precursor indeed appears to be a very luminous, blue
supergiant of type O9 to B2.  This identification of the star's type,
based on color alone, is consistent with that inferred from the
underlying absorption-line features in the SN spectra.

During outburst, as seen in Figure 9b from the 2003 March Palomar
data, the star has developed the observed colors of a late A-type or
early F-type supergiant.  V37, therefore, is following the typical
behavior of a normal LBV during outburst: a ``pseudo-photosphere''
expands in an optically thick stellar wind, increasing the star's $V$
brightness by $sim$2--3 mag (from $V=20.63$ to $V \approx 18$), while
decreasing the apparent stellar temperature to $T_{\rm eff} \approx
8000$ K (although the concept of $T_{\rm eff}$, as applied to LBVs in
eruption, loses its true meaning, and really something more like an
``apparent temperature'' $T_{\rm app}$ is appropriate; see Humphreys
\& Davidson 1994), without a substantial increase in reddening due to
dust formation or change in the bolometric magnitude, $M_{\rm bol}$
(Humphreys \& Davidson 1994).  The formation of extended, dense, dusty
ejecta, such as in the super-outbursts of $\eta$ Car (Humphreys \&
Davidson 1994; Davidson \& Humphreys 1997) and SN 1954J/V12 (Van Dyk
et al.~2005), appears to be relatively exceptional behavior.

At $T_{\rm eff} = 7000$--9000 K, the bolometric correction is zero or
nearly zero (at most, about $-0.2$ mag).  Therefore, for V37 at
maximum light during the outburst, $M_{\rm bol} \approx M_V \approx
-9.8$ mag (assuming $A_V=0.6$ mag and that the 2003 March Palomar $V$
measurement is near maximum, to within $\sim 0.2$ mag).  If we assume
that $M_{\rm bol}$ has remained constant, pre-outburst and at maximum,
then, with the intrinsic pre-outburst color of the star,
$(B-V)_0=-0.13$ mag, and the corresponding $T_{\rm eff}\approx 15500$
K, we can place the star in a HR diagram and estimate the initial mass
of the precursor.

We show the star's location in such a HR diagram in Figure 10.  For
comparison we show the theoretical evolutionary tracks from Lejeune \&
Schaerer (2001), with enhanced mass loss for the most massive stars,
for a 120, 85, 60, and 40 $M_{\sun}$ star.  In Figure 10a we show the
tracks assuming solar metallicity ($Z=0.02$).  It is also possible,
however, that the star may have had abundances different from solar.
SN 2002kg is $119{\farcs}5$ from the galaxy nucleus, which translates
to a galactocentric distance $R\approx 1.9$ kpc.  From a study of the
metallicity gradient in NGC 2403 by Garnett et al.~(1997), the value
of log(O/H)+12 at this distance is 8.6 dex.  With the solar
log(O/H)+12 abundance at 8.8 dex (Grevesse \& Sauval 1998), the
metallicity in this region of the galaxy may actually be somewhat
subsolar; thus, in Figure 10b we show the tracks calculated at the
next lowest metallicity, $Z=0.008$, as well.

What can immediately be seen from this figure, regardless of
metallicity, is that the SN 2002kg precursor, V37, is well off the
zero-age main sequence and is in a more evolved evolutionary phase.
We conclude that the 
initial mass for V37 is $M_{\rm ZAMS} \gtrsim 40\ M_{\sun}$.
Unfortunately, we cannot be more restrictive in our initial mass estimate,
since, due to the uncertainties in the photometry, distance, and reddening of
the star, the termini of the 120, 85, and 60 $M_{\sun}$ tracks are
also consistent with this photometry.  Nonetheless, the HR diagrams
indicate that V37 is indeed a very massive star in the
upper regions of the HR diagram, consistent with the theoretical mass
estimates of stars which go through the LBV phase.

\section{A LBV Nebula Around SN 2002kg/V37?}

From the NOT and ACS/WFC H$\alpha$ images, we can investigate the
gaseous environment of V37, both before and during outburst.  Weis \&
Bomans (2005) have pointed out that in both the {\sl HST\/} images
obtained in the F658N band and ground-based images from 2001, net
H$\alpha$ emission is seen at the position of the star.  However,
Humphreys \& Aaronson (1987) describe a spectrum of the star obtained
in 1985 as consisting of absorption lines and a blue continuum
(consistent with features seen in our own spectra of SN 2002kg/V37),
but with no emission lines visible.  Weis \& Bomans speculate that the
observed H$\alpha$ emission is due to the formation of a LBV nebula,
associated with the current outburst or possibly ejected during an
earlier evolutionary phase (e.g., Lamers et al.~2001).

We confirm that net H$\alpha$ (+ [N~II]) emission is associated
with the star at least as early as 1997, by subtracting the NOT $R$
continuum image from the H$\alpha$ narrow-band image, after first
intensity-scaling the two images (this technique can do a reasonable
job of removing the continuum, when a specific narrow-line bandpass
filter contiguous to H$\alpha$ in wavelength is not available; see
Waller 1990).  We show the result of this subtraction in Figure 11;
there is a very compact (i.e., unresolved and not extended)
H$\alpha$ (+ [N~II]) source at the exact position of the star.
 
We would expect a star with spectral type O9--B2 to have an associated
photoionization (H~II) region.  If we scale the NOT net
H$\alpha$ (+ [N~II]) image by fluxes for the giant H~II
regions S158, S256, and S298 in NGC 2403 (Sivan et al.~1990; Drissen
et al.~1999), we find a flux $F_{{\rm H}\alpha+[{\rm N II}]} \approx
1.1 \times 10^{-15}$ erg cm$^{-2}$ s$^{-1}$ for the nebula
(uncorrected for extinction).  We can correct for extinction using the
relation $A_{{\rm H}\alpha}=0.81A_V$ from Viallefond \& Goss (1986),
and with $A_V=0.6$ mag (assuming the line-of-sight extinction to the
H~II region is the same as that toward the precursor), we find
$1.7 \times 10^{-15}$ erg cm$^{-2}$ s$^{-1}$.  At the distance of NGC
2403 (Freedman et al.~2001) this results in a luminosity $L_{{\rm
H}\alpha+[{\rm N II}]}=2.1 \times 10^{36}$ erg s$^{-1}$.  Using the
relation $N_{\rm Lyc} = 7.25 \times 10^{11} L_{{\rm H}\alpha}$ to
derive the number of ionizing Lyman continuum photons ($1.5 \times
10^{48}$), and using Table II of Panagia (1973), we find that this is
the approximate equivalent of the ionizing flux from a single B0 I
star.  That is, the H$\alpha$ emission seen in 1997 is at least consistent with
a H~II region produced by the ionizing flux of a single massive
blue supergiant --- the outburst precursor, V37.

Subtracting the continuum in the F606W band from the 2004 August
ACS/WFC F658N image, and scaling the net emission image to the fluxes
of the giant H~II regions S256 and S298, results in a flux
$F_{{\rm H}\alpha+[{\rm N II}]} \approx 5.2 \times 10^{-15}$ erg
cm$^{-2}$ s$^{-1}$ (uncorrected for extinction) for SN 2002kg.  This
is approximately a five-fold increase in the H$\alpha$ flux during
outburst in late 2004, relative to the pre-outburst state.  We have
already seen from \S 2.2 that in 2003 January, prior to maximum light,
the star was already in outburst and was emitting $\sim 12$ times more
H$\alpha$ flux than in its quiescent state.  This 2004 August flux
from the {\sl HST\/} image is consistent with the declining fluxes
derived from the integrated spectral line profiles from 2003 through
2004 (see \S 2.2).

The outburst H$\alpha$ flux is likely originating in the dense,
optically thick wind from the star.  But, can we really consider this
a LBV nebula?  The relatively strong [N~II] lines do imply the
presence of circumstellar gas, as does the indication of CNO-processed
material in the wind.  However, a normal LBV eruption does not
generally result in the ejection of a massive shell, or nebula, in the
first place.  If SN 2002kg is similar to S Dor, then we might expect
any nebula to be quite compact.  Based on {\sl HST\/} imaging, Weis
(2003) places an upper limit on the S Dor nebula diameter of 0.25 pc.
Similarly, we find that the source is unresolved ($\lesssim$1.8 WFC1
pixels FWHM) in the 2004 August {\sl HST\/} F658N image which, at the
distance to NGC 2403 (3.13 Mpc; Freedman et al.~2001), implies a
diameter $\lesssim$1.4 pc for the nebula.  If the wind has been
expanding at a constant 370 km s$^{-1}$ between 2002 Oct 26 and 2004
Aug 17, this implies that the wind-blown shell, or nebula, is only
$\sim 0.001$ pc in diameter.  We infer that the detected H$\alpha$
emission is from the ejecta of the recent eruption.

\section{Discussion and Conclusions}

We have shown that SN 2002kg in NGC 2403 is unlike a typical Type II,
core-collapse SN, given both its spectroscopic and photometric
properties.  Although spectroscopically SN 2002kg resembles the SNe
IIn, the ejecta expansion velocity, $\sim$370 km s$^{-1}$, is much
lower than what is observed for the prototypical SNe IIn (e.g., SN
1988Z; Turatto et al.~1993).  Additionally, the star increased only
$\sim 2$ mag in $V$.  We deduce, therefore, that SN 2002kg is, in
fact, not a genuine SN (strictly defined as the explosion of a star at
the end of its life), but instead represents an eruption of a normal,
or ``classical,'' supergiant LBV star, possibly analogous to S Dor.
With $M_{\rm bol} \approx -9.8$ mag and $\log T_{\rm eff}\approx 4.19$
for the precursor in its quiescent state, the star clearly lies on the
``S Dor instability strip'' (Wolf 1989; Humphreys \& Davidson 1994).

We further support this conclusion by showing in Figure 12 the
absolute light curve for SN 2002kg, relative to the light curves for
other SN impostors (SNe 1961V, 1954J/V12 1997bs and 2000ch) in a
similar vein as Van Dyk et al.~(2000; their Figure 6).  Although the
bandpasses through which these light curves were generated differ from
one object to the other, the relative similarities in luminosity,
energy, and variability argue strongly that SN 2002kg is also a SN
impostor.

SN 2002kg appears from its light curve to be the least energetic
example of a SN impostor known so far.  SNe 1954J/V12, 1961V, and
$\eta$ Car may represent the high-energy end of a pre-SN
super-outburst phase, with SN 1997bs and SN 2000ch somewhere in
between.  The lower energetics can also be seen in a comparison
(Fig. 13) of the early-time spectra of SNe 2002kg, 1997bs, and 2000ch.
The overall strength of the Balmer-line profiles in SN 2002kg is lower
than in the other two SNe. Moreover, the bases of the profiles in SN
2002kg are not nearly as broad as in SNe 1997bs and 2000ch, consistent
with the lower ejecta velocities in SN 2002kg. (SN 2000ch also shows
more of a true P-Cygni profile, with shallow blue absorption troughs
in the Balmer lines.)  SN 2002kg is clearly less energetic than SN
1954J/V12 in the same host galaxy, NGC 2403; this is reflected both in
the lower expansion velocity ($\sim 370$ km s$^{-1}$, versus $\sim
700$ km s$^{-1}$ at age $\sim$50 yr; Van Dyk et al.~2005), and fainter
apparent brightness ($m_V [{\rm max}] \approx 18.3$ mag versus 16.1
mag; Humphreys \& Davidson 1994).

The precursor of SN 2002kg is the irregular blue variable star V37 in
NGC 2403, identified by Humphreys \& Aaronson (1987) and Humphreys \&
Davidson (1994) as a LBV.  We have also directly placed constraints on
the initial mass of this precursor to be $\gtrsim 40\ M_{\sun}$.  As a
result of the outburst, the star became redder and attained a nearly
constant color (the $B-V$ color remained relatively unchanged for
nearly 1.5 yr; the overall constant shape of the spectral continua
further supports this), as the pseudo-photosphere of the optically
thick envelope expands.  The absolute flux in the H$\alpha$ line
decreased between early 2003 and late 2004; this is also reflected in
the $B-R$ color of the star (0.41 mag in 2003 March and 0.26 mag in
2004 August/September).  The constant color argues against any dust
formation event for SN 2002kg/V37, such as is evident in the
Homunculus around $\eta$ Car and in SN 1954J/V12 (Van Dyk et
al.~2005), although dust might form later for the SN 2002kg/V37 ejecta.

SN 2002kg/V37 is one of a growing number of objects
which``impersonate'' SNe IIn.  These impostors exhibit a broad range
of properties and actually represent a pre-SN evolutionary phase of
very massive stars. It is imperative to continue monitoring the star
to detail its further evolution.  Other recent, less well-studied
cases, such as SN 1999bw in NGC 3198, SN 2001ac in NGC 3504, and SN
2003gm in NGC 5334, also require detailed study, to provide valuable
additional examples of this rare phenomenon and improve our
understanding of massive-star evolution.

\acknowledgements

We acknowledge the assistance of the Lick, MMT, and Keck Observatory
staffs.  R.M.H. thanks R. Mark Wagner and Michael Koppelman for their
help, respectively, with the MMT observations and data reductions.  We
are very grateful to Soeren Larsen for assistance in the calibration
of his images.  We thank Nolan Walborn to pointing us to the digital
spectrum for AG Car, and Phil Massey for sending us the digital
spectrum of S Dor.  This research is partially based on data from the Isaac
Newton Group Archive.  This publication makes use of data products from
the Two Micron All Sky Survey, which is a joint project of the
University of Massachusetts and the IPAC/California Institute of
Technology, funded by NASA and NSF.  This research also utilizes the
NASA/IPAC Extragalactic Database (NED) which is operated by the Jet
Propulsion Laboratory, California Institute of Technology, under
contract with NASA.  The work of A.V.F.'s group at UC Berkeley is
supported by NSF grant AST-0307894, as well as by NASA grants AR-9953,
AR-10297, AR-10690, GO-10182, and GO-10272 from the Space Telescope
Science Institute, which is operated by AURA, Inc., under NASA
contract NAS5-26555. A.V.F. is grateful for a Miller Research
Professorship at UC Berkeley, during which part of this work was
completed.  KAIT was made possible by generous donations from Sun
Microsystems, Inc., the Hewlett-Packard Company, AutoScope
Corporation, Lick Observatory, the National Science Foundation, the
University of California, and the Sylvia and Jim Katzman Foundation.
The W. M. Keck Observatory is operated as a scientific partnership
among the California Institute of Technology, the University of
California, and NASA; the observatory was made possible by the
generous financial support of the W. M. Keck Foundation.


\begin{deluxetable}{lcc}
\tablenum{1}
\tablewidth{4.5truein}
\tablecolumns{3}
\tablecaption{KAIT Photometry of SN 2002kg}
\tablehead{
\colhead{UT date} &\colhead{JD$-$2450000 } & 
\colhead{$R$ (mag)}}
\startdata
1998 Nov 13\phantom{.00} & 1130\phantom{.00} & \llap{$\ga$}19.5\phantom{(0.1)}  \\
2002 Oct 26.56 & 2574.06 & 18.4(0.1) \\
2002 Nov 01.53 & 2580.03 & 18.3(0.1) \\
2002 Nov 12.54 & 2591.04 & 18.1(0.2) \\
2002 Nov 16.44 & 2594.94 & 18.4(0.1) \\
2002 Nov 20.40 & 2598.90 & 18.4(0.2) \\
2002 Dec 02.49 & 2610.99 & 18.0(0.1) \\
2003 Jan 08.40 & 2647.90 & 17.9(0.1) \\
2003 Jan 16.29 & 2655.79 & 18.0(0.1) \\
2003 Jan 26.34 & 2665.84 & 18.2(0.1) \\
2003 Jan 31.27 & 2670.77 & 18.2(0.1) \\
2003 Feb 08.26 & 2678.76 & 18.1(0.1) \\
2003 Feb 23.23 & 2693.73 & 18.1(0.1) \\
2003 Mar 03.20 & 2701.70 & 18.0(0.1) \\
2003 Mar 08.17 & 2706.67 & 18.1(0.1) \\
2003 Mar 13.21 & 2711.71 & 17.8(0.1) \\
2003 Mar 25.17 & 2723.67 & 18.0(0.1) \\
2003 Oct 11.49 & 2923.99 & 18.1(0.1) \\
2003 Oct 18.54 & 2931.04 & 18.3(0.1) \\
2003 Oct 26.54 & 2939.04 & 18.0(0.1) \\
2003 Nov 10.53 & 2954.03 & 18.1(0.1) \\
2003 Nov 23.48 & 2966.98 & 18.1(0.1) \\
2004 Feb 01.32 & 3036.82 & 18.5(0.1) \\
2004 Feb 11.26 & 3046.76 & 18.8(0.1) \\
2004 Mar 03.19 & 3068.69 & 19.1(0.3) \\
2004 Mar 08.13 & 3073.63 & 19.4(0.2) \\
2004 Mar 17.16 & 3082.66 & 18.7(0.1) \\
2004 Apr 04.20 & 3099.70 & 18.4(0.1) \\
\enddata
\tablenotetext{}{Uncertainties in the magnitudes are given in parentheses.}
\end{deluxetable}

\clearpage

\begin{deluxetable}{lcccccc}
\tablenum{2}
\tablewidth{6.5truein}
\tablecolumns{7}
\tablecaption{SN 2002kg Photometry from Other Sources}
\tablehead{
\colhead{UT date} & \colhead{$U$} & \colhead{$B$} & \colhead{$V$}
& \colhead{$R$} & \colhead{$I$} & \colhead{Source\tablenotemark{a}} \\
\colhead{} & \colhead{(mag)} & \colhead{(mag)} & \colhead{(mag)}
& \colhead{(mag)} & \colhead{(mag)} & \colhead{}
}
\startdata
1987 Feb 04 & \nodata & $\ga 20.3$ & $\ga 20.4$ & $\ga20.1$ & \nodata & JKT \\
1992 Mar 22 & \nodata & $\ga 16.0$ & $\ga 15.8$ & $\ga 15.7$ & \nodata & INT \\
1997 Oct 13 & 19.59(04) & 20.55(03) & 20.63(04) & 20.54(02) & 20.69(03) & NOT \\
1998 Oct 05 & \nodata & $\ga 20.5$ & \nodata & $\ga 20.4$ & \nodata & JKT \\
1999 Apr 25 & \nodata & $\ga 20.1$ & \nodata & $\ga 20.1$ & \nodata & JKT \\
2003 Mar 26 & \nodata & 18.45(03) & 18.32(04) & 18.04(02) & 17.97(02) & P60 \\
2004 Aug 17\tablenotemark{b} & \nodata & 19.58(05) & 19.39(06)  & \nodata &19.21(06) & {\sl HST} \\
2004 Sep 21\tablenotemark{b} & \nodata & 19.41(07)& \nodata & 19.13(04) & \nodata & {\sl HST} \\
\enddata
\tablenotetext{}{Note: uncertainties of hundredths of a magnitude are indicated in parentheses.}
\tablenotetext{a}{JKT=Johannes Kepler Telescope; INT=Isaac Newton Telescope; NOT=Nordic Optical Telescope; 
P60=Palomar 1.5-m telescope; {\sl HST}={\sl Hubble Space Telescope}.}
\tablenotetext{b}{Transformed from the flight system magnitudes via synthetic photometry of normal stars with SYNPHOT; see text.}
\end{deluxetable}


\begin{figure}
\figurenum{1}
\caption{SN 2002kg (indicated by tick marks) in NGC 2403 in an
unfiltered ($\sim R$) image obtained with the Katzman Automatic Imaging
Telescope (KAIT) 0.76-m telescope on 2002 October 26.  The image is
$6{\farcm}6 \times 6{\farcm}6$ in size.  North is up, and east is to
the left.}
\end{figure}

\clearpage

\begin{figure}
\figurenum{2}
\plotone{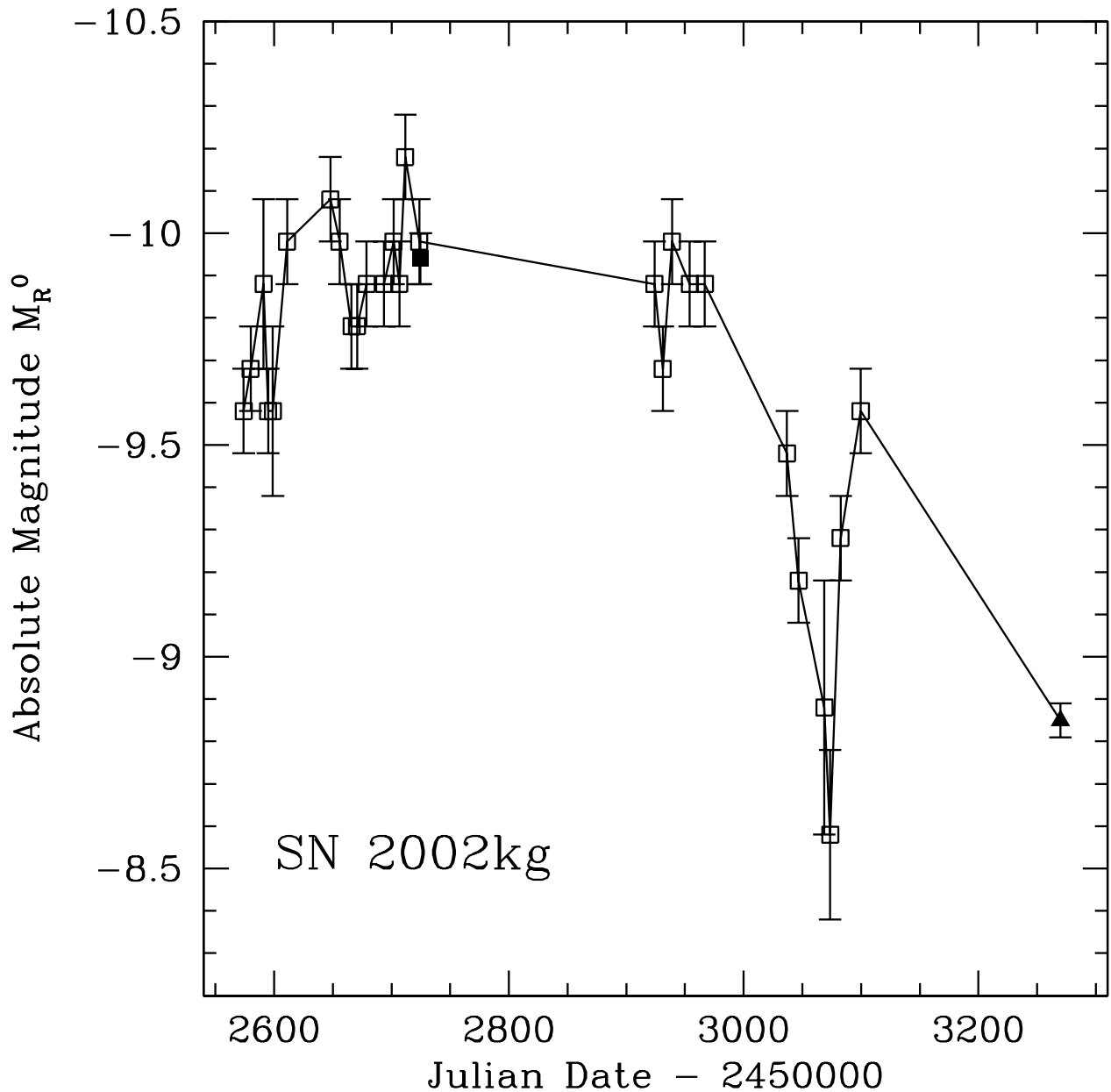}
\caption{The unfiltered ($\sim R$-band) light curve ({\it open
squares}) of SN 2002kg obtained using KAIT.  (See Table 1; see text
regarding the photometric zero point and systematic errors in the
light curve.)  Also shown are the $R$-band data for the SN obtained
with the Palomar 1.5-m telescope on 2003 March 26.15 ({\it filled
square}) and with the {\sl HST\/} ACS on 2003 September 21.07 ({\it
filled triangle}).  The photometry has been corrected for extinction
(see text) and for the true distance modulus of the host galaxy, NGC
2403 (Freedman et al.~2001).}
\end{figure}

\clearpage

\begin{figure}
\figurenum{3}
\caption{SN 2002kg (indicated by the arrow) in a $V$-band image obtained at 
the Palomar 1.5-m telescope on 2003 March 26.15.}
\end{figure}


\begin{figure}
\figurenum{4}
\caption{SN 2002kg (indicated by the arrow) in a F606W image obtained
with the ACS/WFC on-board {\sl HST\/} on 2004 August 17.11.  This
figure is to approximately the same scale and orientation as Figure 3.
Note that the SN field is near the edge of the ACS image.}
\end{figure}

\begin{figure}
\figurenum{5}
\plotone{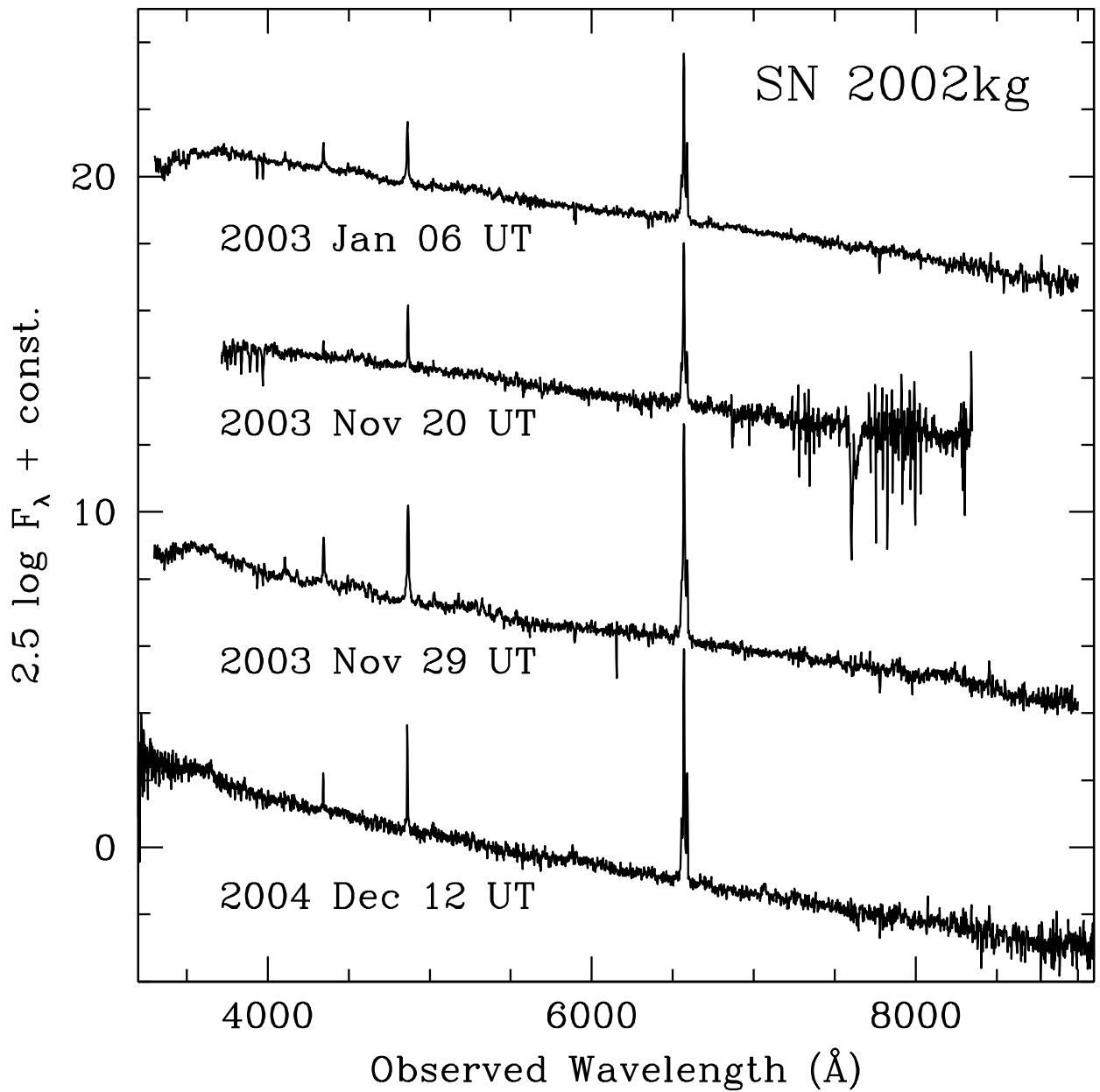}
\caption{Spectra of SN 2002kg obtained at the 10-m Keck telescope
(2003 Jan 6, 2003 Nov 29, and 2004 Dec 12) and the 6.5-m
MMT (2003 Nov 20).}
\end{figure}

\clearpage

\begin{figure}
\figurenum{6}
\plotone{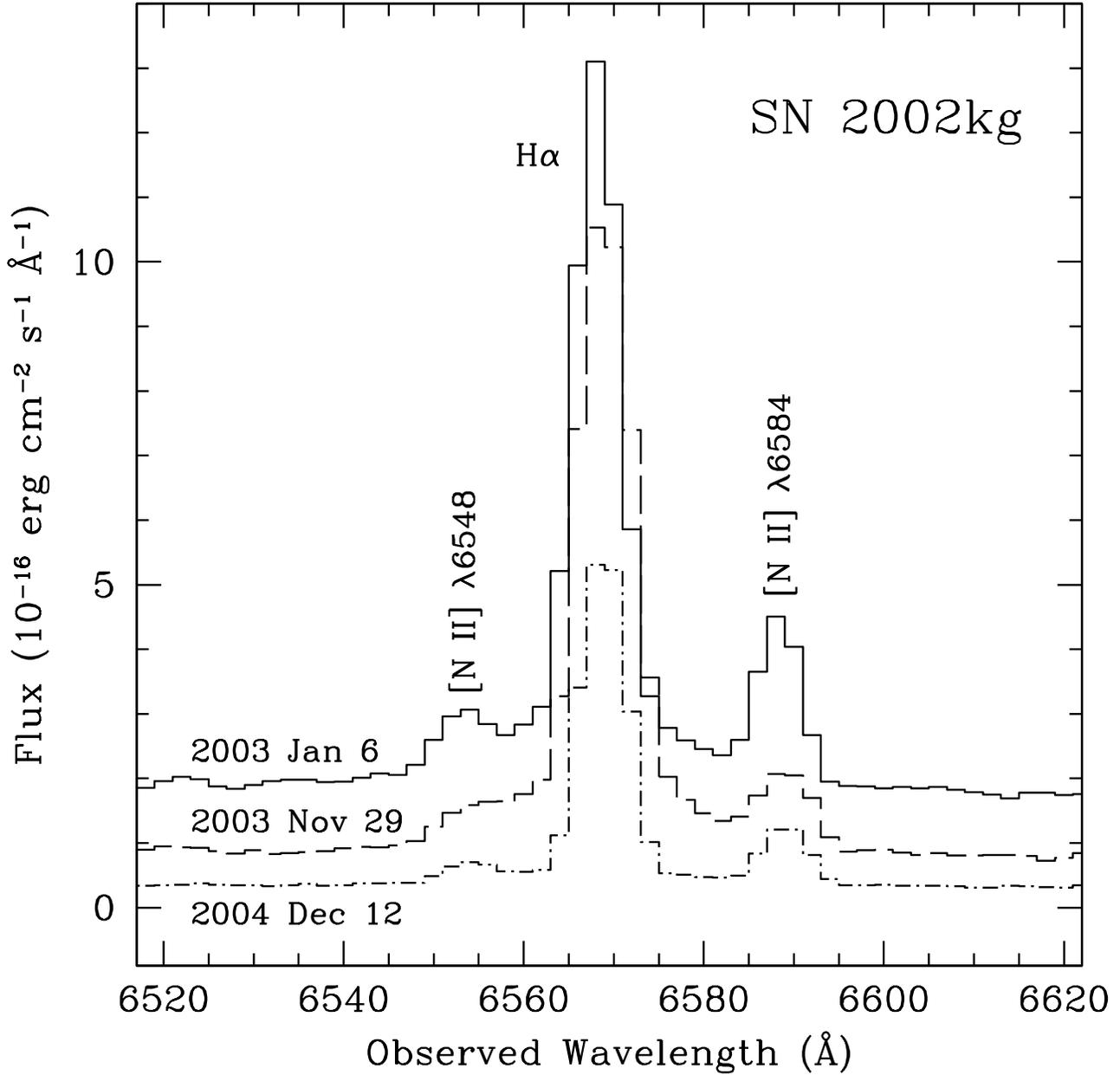}
\caption{A comparison of the spectra shown in Figure 5 in the
wavelength region containing H$\alpha$ and the [N II] lines.  The
narrow emission component of H$\alpha$ is unresolved, and the broad
component remains relatively constant in width throughout 2003 and
2004. The prominent, unresolved [N II] $\lambda\lambda$6548, 6584 \AA\
emission lines suggest the presence of N-enhanced circumstellar
material.}
\end{figure}

\clearpage

\begin{figure}
\figurenum{7}
\plotone{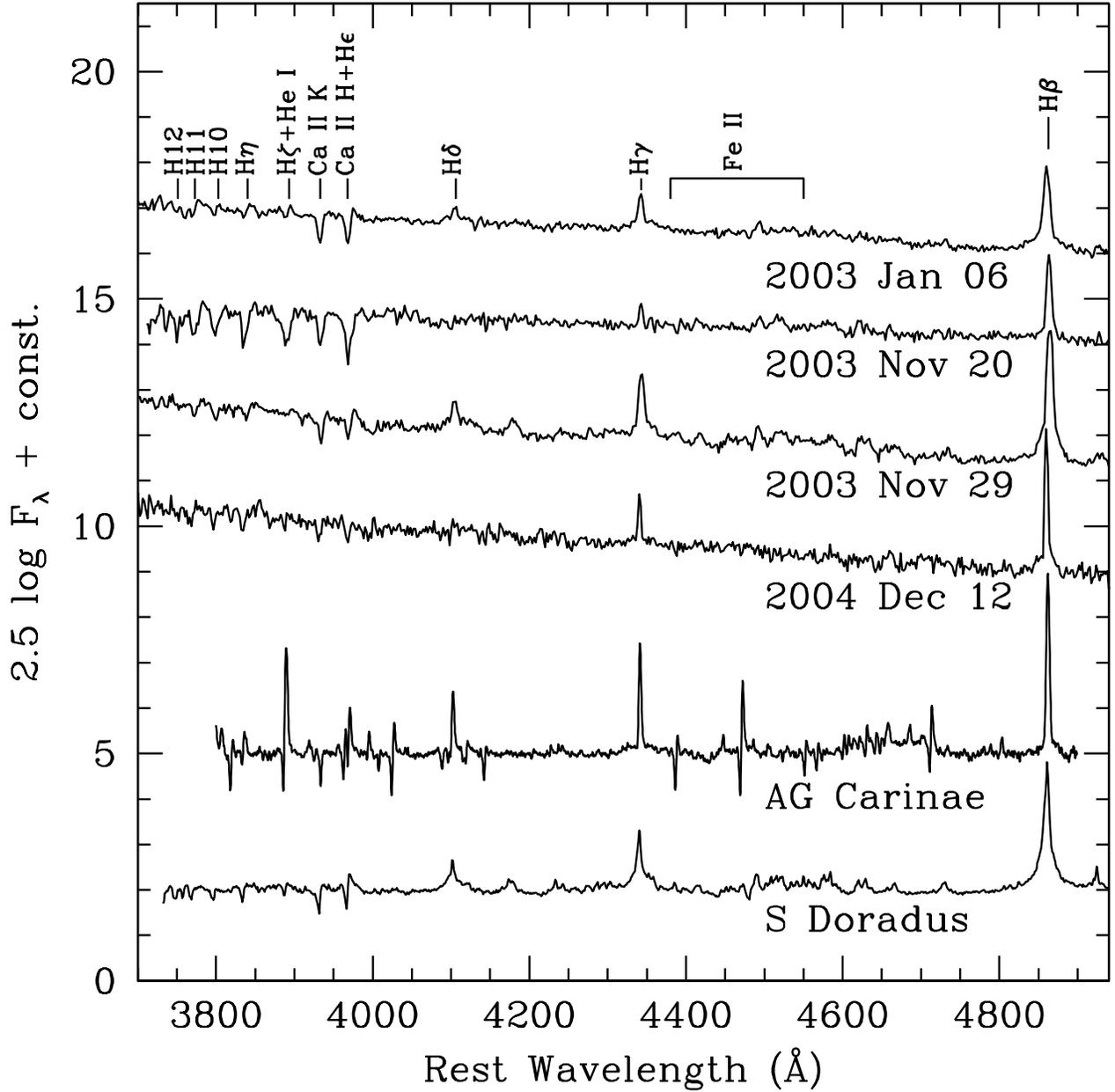}
\caption{Expanded views of the spectra of SN 2002kg shown in Figure 5 (redshift-corrected
assuming $z=0.000437$),
in the blue wavelength region.  Also shown for comparison are spectra
of the LBVs AG Car (Walborn \& Fitzpatrick 2000) and S Dor (Massey
2000; redshift-corrected assuming $z=0.000927$).  Various lines are labelled.}
\end{figure}


\begin{figure}
\figurenum{8}
\caption{The precursor of SN 2002kg (indicated by the arrow), as seen
in a $V$-band image of the host galaxy, NGC 2403, obtained with the
2.6-m Nordic Optical Telescope (NOT) on 1997 Oct 13 (Larsen \&
Richtler 1999).  This figure is to nearly the same scale and
orientation as in Figure 3.  The location of the precursor is
coincident with the irregular blue variable V37, reported by Tammann
\& Sandage (1968).}
\end{figure}

\clearpage

\begin{figure}
\figurenum{9}
\plotone{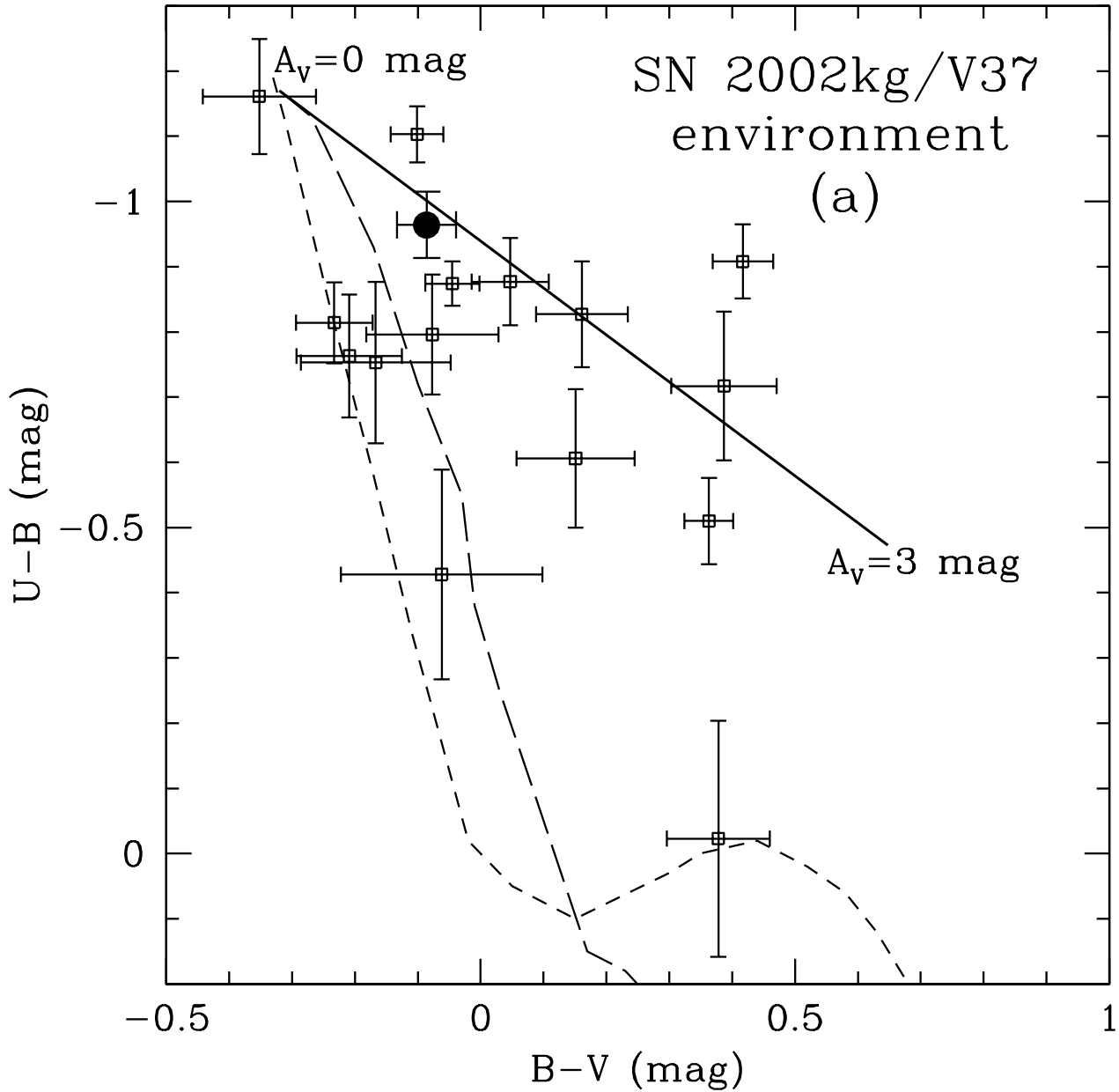}
\caption{The ({\it a}) ($U-B$, $B-V$) and ({\it b}) ($B-V$, $V-I$)
color-color diagrams of SN 2002kg and its stellar environment (within
$\sim 500$ pc).  The colors of the precursor star ({\it filled
circle}) measured from the NOT images are shown in both diagrams.  The
{\it open squares\/} represent the colors of the stars in the
environment measured from the NOT images for panel ({\it a}) and the
{\sl HST\/} ACS/WFC images for panel ({\it b}) (after transformation
from flight system to standard Johnson-Cousins colors).  Also shown
are the loci for supergiant ({\it long-dashed line}) and main sequence
({\it short-dashed line}) stars, as well as the reddening vector ({\it
solid line}) from $A_V=0$ to $A_V=3$ mag, assuming the Cardelli et
al.~(1989) reddening law. From these diagrams, we infer $A_V = 0.6
\pm 0.2$ mag for the precursor star.  In ({\it b}) we show the colors
of the SN during outburst from the 2003 Palomar data ({\it filled
square}) and from the 2004 {\sl HST}/ACS data ({\it filled triangle};
the {\it open triangle\/} represents the SN colors after an assumed
correction for contamination by the strong H$\alpha$ emission in the
F606W band; a {\it dotted line\/} connects the two symbols).}
\end{figure}

\clearpage

\begin{figure}
\figurenum{9}
\plotone{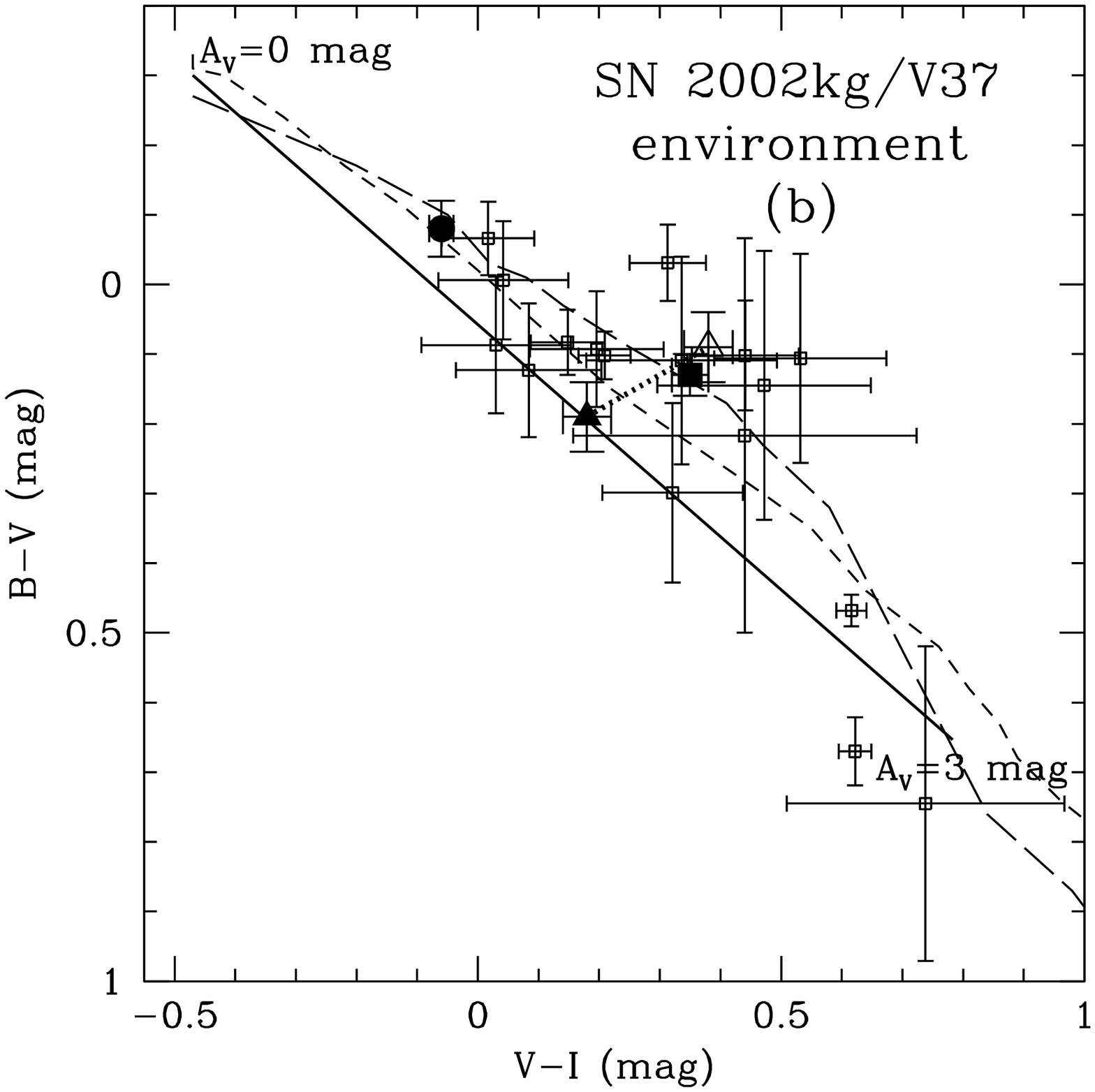}
\caption{(Continued.) }
\end{figure}

\clearpage

\begin{figure}
\figurenum{10}
\plotone{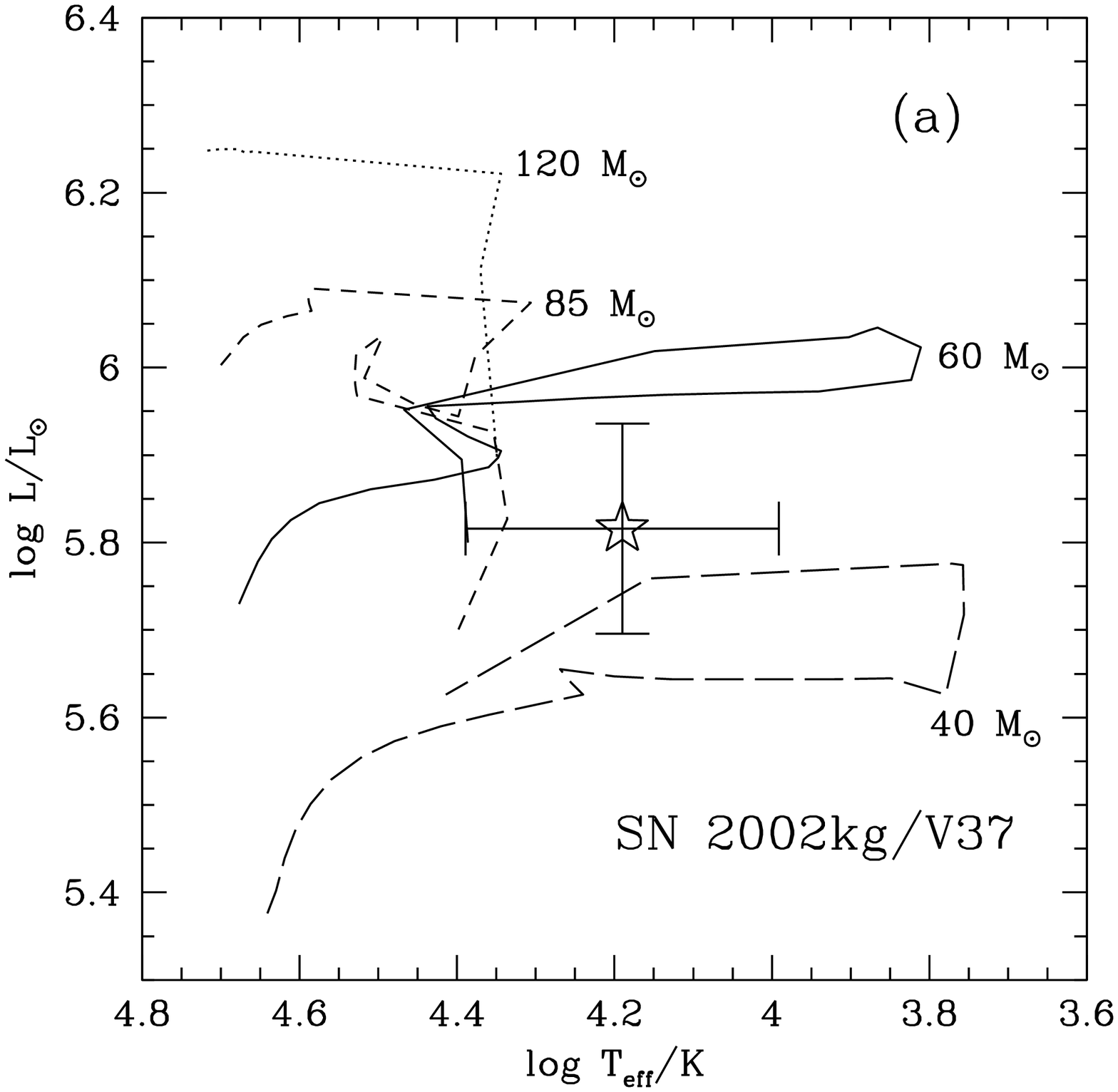}
\caption{HR diagrams, showing the locus for the precursor ({\it
five-pointed star}), after correction for reddening and the true
distance modulus of the host galaxy, NGC 2403, and assuming that
$M_{\rm bol}$ has remained constant before outburst and at maximum.
Also shown for comparison are the model stellar evolutionary tracks
from Lejeune \& Schaerer (2001), with enhanced mass loss, for 120
$M_{\sun}$ ({\it dotted line}), 85 $M_{\sun}$ ({\it short-dashed
line}), 60 $M_{\sun}$ ({\it solid line}), and 40 $M_{\sun}$ ({\it
long-dashed line}).  Panel (a) shows the tracks for solar metallicity
($Z=0.02$), and panel (b) shows the tracks for sub-solar metallicity
($Z=0.008$).}
\end{figure}

\clearpage

\begin{figure}
\figurenum{10}
\plotone{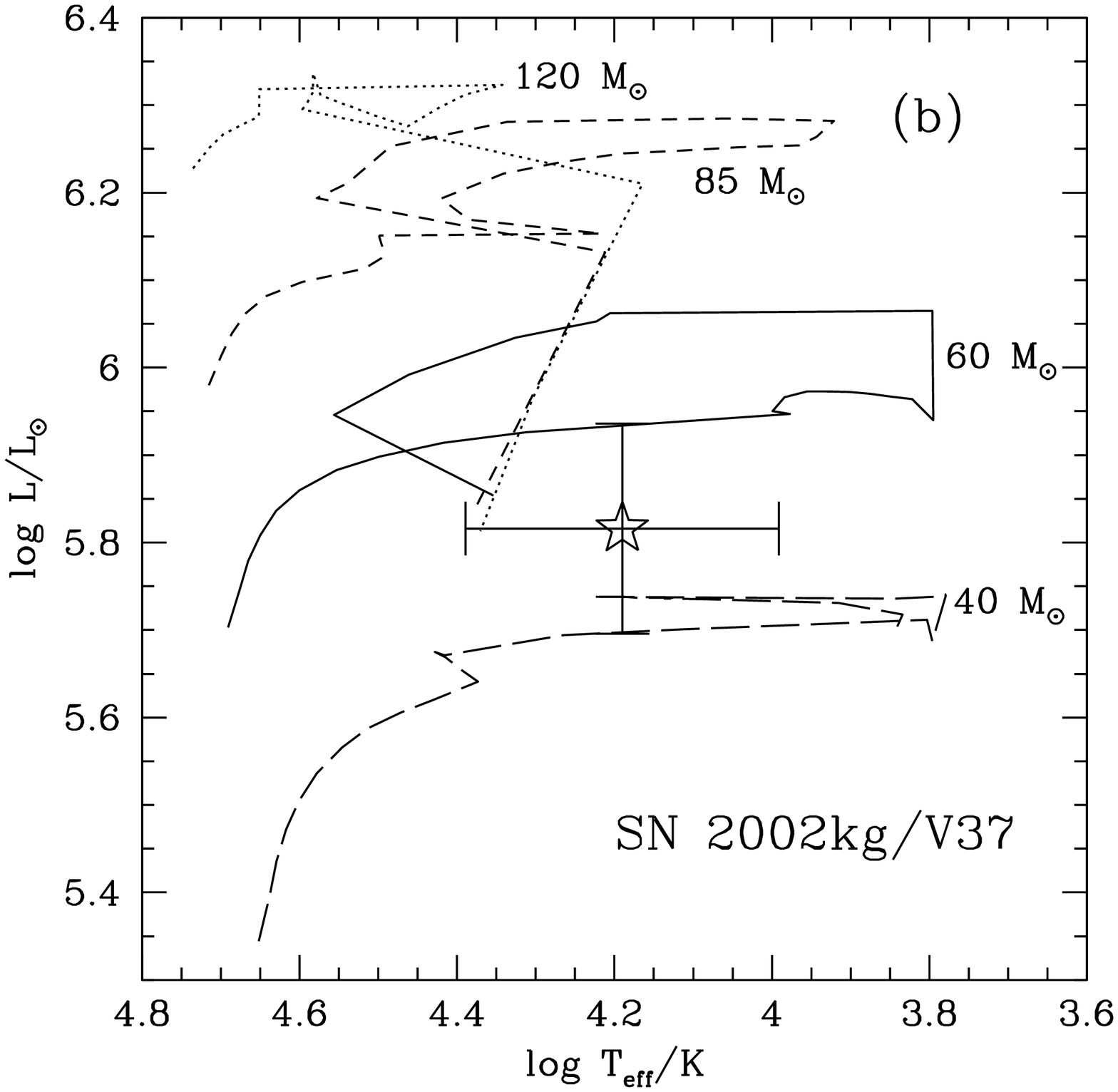}
\caption{(Continued.)} 
\end{figure}

\clearpage

\begin{figure}
\figurenum{11}
\caption{The continuum-subtracted H$\alpha$ image of the host galaxy,
NGC 2403, obtained with the 2.6-m Nordic Optical Telescope (NOT) on
1997 Oct 13 (Larsen \& Richtler 1999), showing net emission
surrounding the precursor of SN 2002kg (indicated by the arrow).  This
figure is shown to the same scale and orientation as Figure 8.  The
nebula in this image is spatially unresolved, with diameter $\lesssim
14$ pc (FWHM).  The H$\alpha$ luminosity is consistent with
photoionization by a single blue supergiant star.  Artifacts, due to
incomplete continuum subtraction, are also seen in the figure.}
\end{figure}


\begin{figure}
\figurenum{12}
\plotone{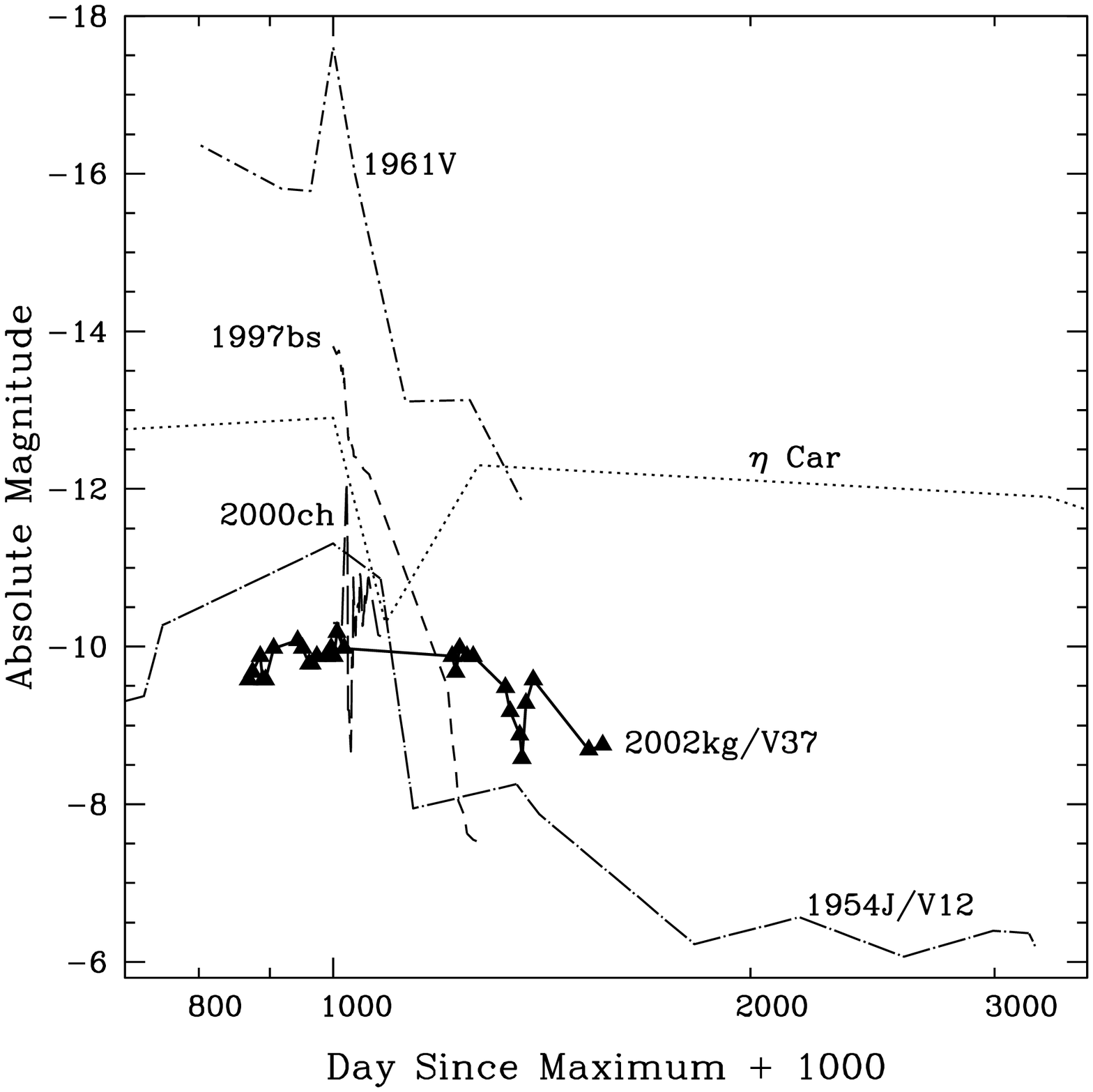}
\caption{A comparison of the absolute light curve for SN 2002kg in NGC
2403 ({\it filled triangles\/} and {\it solid line}; also see Figure
2) with that of $\eta$ Car ({\it dotted line}), SN 1954J in NGC 2403
({\it long-dashed, dotted line}), SN 1961V in NGC 1058 ({\it
short-dashed, dotted line}; adapted from Humphreys, Davidson, \& Smith
1999), SN 1997bs in M66 ({\it short-dashed line}; Van Dyk et
al.~2000), and SN 2000ch in NGC 3432 ({\it long-dashed line}; Wagner
et al.~2004).  We note that the
bandpasses through which each of these light curves were generated differed 
from one object to the other.}
\end{figure}

\clearpage

\begin{figure}
\figurenum{13}
\plotone{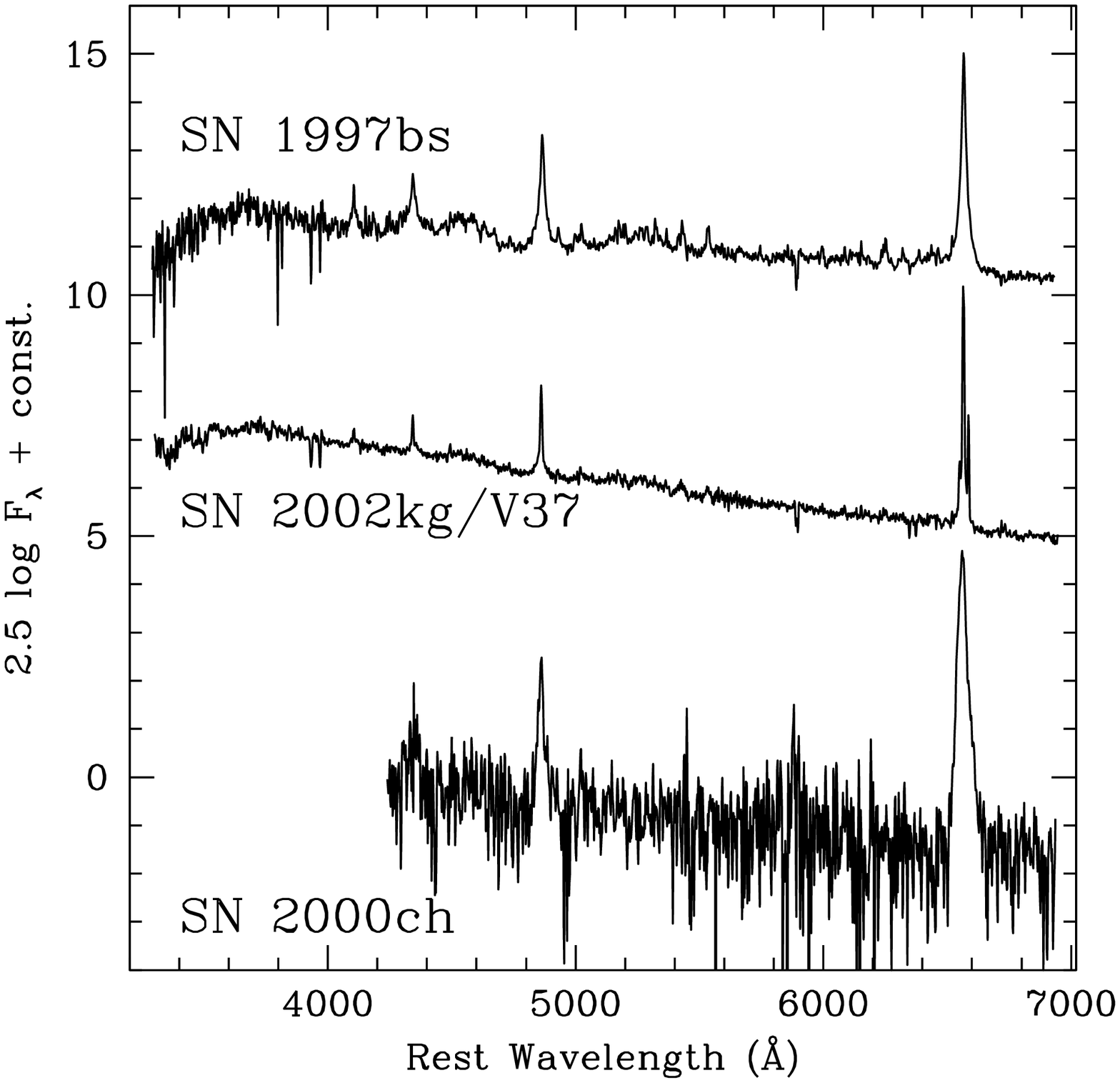}

\caption{A comparison of the early-time optical spectrum of SN 2002kg
(see Figure 5; redshift-corrected assuming $z=0.000437$) with those of 
SN 1997bs in M66 (Van Dyk et al.~2000; redshift-corrected assuming
$z=0.002425$)
and SN 2000ch in NGC 3432 (Wagner et al.~2004; redshift-corrected 
assuming $z=0.002055$).}

\end{figure}

\end{document}